\documentclass[aps,prl,twocolumn,showpacs,superscriptaddress,floatfix,reprint]{revtex4-1}
\usepackage{pstricks,pst-node,pst-text,pst-3d,pst-plot}
\usepackage{graphicx}
\usepackage{bm}
\usepackage{amssymb}  

\usepackage[english]{babel}
\usepackage{eepic} 

\def\reff#1{(\ref{#1})}
\newcommand{\be}{\begin{equation}}
\newcommand{\ee}{\end{equation}}

\def\spose#1{\hbox to 0pt{#1\hss}}
\def\ltapprox{\mathrel{\spose{\lower 3pt\hbox{$\mathchar"218$}}
 \raise 2.0pt\hbox{$\mathchar"13C$}}}
\def\gtapprox{\mathrel{\spose{\lower 3pt\hbox{$\mathchar"218$}}
 \raise 2.0pt\hbox{$\mathchar"13E$}}}

\newcommand{\x}{\hphantom{0}}

\newsavebox{\fancyplusb}

\savebox{\fancyplusb}(2,2){
\thinlines
\put(0,-1){\line(0,1){2}}
\put(-1,-0.5){\line(0,1){1}}
\put(-1,0){\line(1,0){2}}
\put(1,-0.5){\line(0,1){1}}
\put(-0.5,-1){\line(1,0){1}}
\put(-0.5,1){\line(1,0){1}}
}

\begin{document}

\title{Two-dimensional Potts antiferromagnets with a phase transition
       at arbitrarily large $\bm{q}$}


\author{Yuan Huang}
\email{huangy22@mail.ustc.edu.cn}
\author{Kun Chen}
\email{chenkun@mail.ustc.edu.cn}
\author{Youjin Deng}
\email{yjdeng@ustc.edu.cn}
\affiliation{Hefei National Laboratory for Physical Sciences at Microscale
   and Department of Modern Physics,
   University of Science and Technology of China,
   Hefei, Anhui 230026, China}
\author{Jesper Lykke Jacobsen}
\email{jacobsen@lpt.ens.fr}
\affiliation{Laboratoire de Physique Th\'eorique,
   \'Ecole Normale Sup\'erieure, 24 rue Lhomond, 75231 Paris, France}
\affiliation{Universit\'e Pierre et Marie Curie,
   4 place Jussieu, 75252 Paris, France}
\author{Roman~Koteck\'y}
\email{R.Kotecky@warwick.ac.uk}
\affiliation{Center for Theoretical Study, Charles University, Prague,
  Czech Republic}
\affiliation{Mathematics Institute, University of Warwick, Coventry CV4 7AL, UK}
\author{Jes\'us Salas}
\email{jsalas@math.uc3m.es}
\affiliation{
  Escuela Polit\'ecnica Superior,  
  Universidad Carlos III de Madrid,
  28911 Legan\'es, Spain}
\affiliation{
  Grupo de Teor\'{\i}as de Campos y F\'{\i}sica Estad\'{\i}stica,
  Gregorio Mill\'an Institute,
  Universidad Carlos III de Madrid,
  Unidad asociada al IEM-CSIC, Madrid, Spain}
\author{Alan D. Sokal}
\email{sokal@nyu.edu}
\affiliation{Department of Physics, New York University,
      4 Washington Place, New York, NY 10003, USA}
\affiliation{Department of Mathematics, University College London,
      London WC1E 6BT, UK}
\author{Jan M. Swart}
\email{swart@utia.cas.cz}
\affiliation{Institute of Information Theory and Automation (\' UTIA),
      18208 Prague 8, Czech Republic}

\date{January 8, 2013}

\begin{abstract}
We exhibit infinite families of two-dimensional lattices
(some of which are triangulations or quadrangulations of the plane)
on which the $q$-state Potts antiferromagnet has
a finite-temperature phase transition at arbitrarily large values of $q$.
This unexpected result is proven rigorously
by using a Peierls argument
to measure the entropic advantage of sublattice long-range order.
Additional numerical data are obtained using transfer matrices,
Monte Carlo simulation,
and a high-precision graph-theoretic method.
\end{abstract}



\maketitle

The $q$-state Potts model \cite{Potts_52,Wu_82+84}
plays an important role in the theory of critical phenomena,
especially in two dimensions (2D) \cite{Baxter_book,Nienhuis_84,DiFrancesco_97},
and has applications to various condensed-matter systems \cite{Wu_82+84}.
Ferromagnetic Potts models are by now fairly well understood,
thanks to universality;
but the behavior of antiferromagnetic Potts models
depends strongly on the microscopic lattice structure,
so that many basic questions
about the phase diagram and critical exponents  
must be investigated case-by-case.
In this article we prove
the unexpected existence of phase transitions for
some 2D $q$-state Potts antiferromagnets at arbitrarily large values of $q$.

For Potts antiferromagnets
one expects that for each lattice ${\cal L}$ there
is a value $q_c({\cal L})$ [possibly noninteger]
such that for $q > q_c({\cal L})$  the model has exponential decay~of
correlations at all temperatures including zero,
while for $q = q_c({\cal L})$
there is a zero-temperature critical point.
The first task, for any lattice, is thus to determine $q_c$.

Some 2D antiferromagnetic models at zero temperature
can be mapped
exactly
onto a ``height''
model
\cite{Salas_98,Jacobsen_09}.
Since the height model must either be in a ``smooth'' (ordered)
or ``rough'' (massless) phase,
the corresponding zero-temperature spin model must either be
ordered or critical, never disordered.
Until now it has seemed that the most common case is criticality
\cite{height_rep_exceptions}.

In particular, when the $q$-state zero-temperature Potts antiferromagnet (AF)
on a
2D lattice ${\cal L}$ admits a height representation,
one ordinarily expects that $q = q_c({\cal L})$.
This prediction is confirmed in most heretofore-studied cases:
3-state square-lattice \cite{Nijs_82,Kolafa_84,Burton_Henley_97,Salas_98},
3-state kagome \cite{Huse_92,Kondev_96},
4-state triangular \cite{Moore_00},
and 4-state on the line graph
of the square lattice
\cite{Kondev_95,Kondev_96}.
Until recently the only known exception was the
triangular Ising AF \cite{note_TRI_q=2}.

Koteck\'y, Salas and Sokal (KSS) \cite{Kotecky-Salas-Sokal}
observed that the height mapping employed for the
3-state Potts AF on the square lattice \cite{Salas_98}
carries over unchanged to any plane quadrangulation;
and Moore and Newman \cite{Moore_00} observed
that the height mapping employed for the
4-state Potts AF on the triangular lattice
carries over unchanged to any Eulerian plane triangulation
(a graph is called Eulerian if all vertices have even degree).
One therefore expects naively that $q_c = 3$
for every (periodic) plane quadrangulation,
and that $q_c = 4$
for every (periodic) Eulerian plane triangulation.

Surprisingly, these predictions are {\em false}\/!
KSS \cite{Kotecky-Salas-Sokal}
proved rigorously that the 3-state
AF on the diced lattice (which is a quadrangulation)
has a phase transition at finite temperature
(see also \cite{Kotecky-Sokal-Swart});
numerical estimates from transfer matrices yield
$q_c({\rm diced}) \approx 3.45$ \cite{Jacobsen-Salas_unpub}.
Likewise, we recently \cite{union-jack} provided analytic arguments
(falling short, however, of a rigorous proof)
that on any Eulerian plane triangulation in which one sublattice
consists
entirely
of vertices of degree 4,
the 4-state AF has a
phase
transition at finite temperature,
so that $q_c > 4$.
We also presented transfer-matrix and Monte Carlo data
confirming these predictions for
the union-jack
and bisected hexagonal
lattices, leading to the estimates $q_c({\rm UJ}) \approx 4.33$
and $q_c({\rm BH}) \approx 5.40$.

These results suggest the obvious question:  How large can $q_c$ be
on a plane quadrangulation (resp.\ Eulerian plane triangulation)?
The answers are clearly larger than 3 or 4, respectively ---
but how much larger?

In this article we shall give a rigorous proof
of the unexpected answer:
we exhibit infinite classes of plane quadrangulations
and Eulerian plane triangulations on which $q_c$ can take
{\em arbitrarily large}\/ values.
We shall also complement this rigorous proof with detailed quantitative data
from transfer matrices, Monte Carlo simulations,
and a powerful graph-theoretic approach
developed recently by Jacobsen and Scullard \cite{JS1}.

The models studied here provide new examples of
{\em entropically-driven long-range order}\/
\cite{Kotecky-Salas-Sokal,union-jack,Kotecky-Sokal-Swart,Chen_11}:
the ferromagnetic ordering of spins on one sublattice
is favored because it increases the freedom of choice of spins
on the other sublattice(s).
But though this idea is intuitively appealing,
it is usually difficult to determine quantitatively, in any specific case,
whether the entropic penalty for interfaces between domains of
differently-ordered spins on the first sublattice
is large enough to produce long-range order.
Moreover, one expects that this penalty decreases with increasing $q$.
In the examples given here, by contrast, we are able to prove that
the penalty can be made arbitrarily strong
and hence operative at arbitrarily large $q$.

\begin{figure}
  \centering
  \begin{tabular}{ccc}
  \includegraphics[width=80pt]{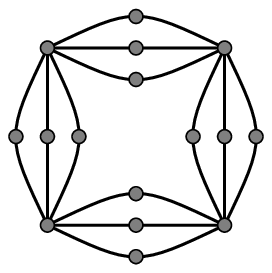} & \makebox[0.7cm]{} & 
  \includegraphics[width=80pt]{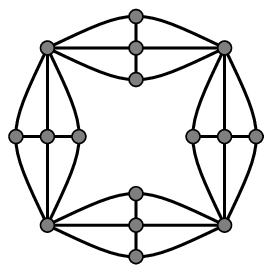} \\
  (a) & \makebox[0.7cm]{} & (b) 
  \end{tabular}
  \caption{%
      Unit cells of the lattices $G_n$ (a) and $H_n$ (b) for $n=3$.
   }
  \label{fig1}
\end{figure}

\paragraph{The lattices $G_n$ and $H_n$.}
Let $G_n$ be obtained from the square (SQ) lattice
by replacing each edge with $n$~two-edge paths in parallel;
and let $H_n$ be obtained from $G_n$ by connecting each group
of $n$~``new'' vertices with an $(n-1)$-edge path
(see Fig.~\ref{fig1}).
Resumming over the spins on the ``new'' vertices \cite{Sokal_bcc2005},
it is easy to show that the $q$-state Potts model on $G_n$ or $H_n$
with nearest-neighbor coupling $v = e^J - 1$
is equivalent to a SQ-lattice Potts model
with a suitable coupling $v_{\rm eff}(q,v)$ \cite{Kotecky_85};
moreover, for $q > 2$ (resp.\ $q > 3$)
an AF model ($-1 \le v \le 0$) on $G_n$ (resp.\ $H_n$)
maps onto a ferromagnetic model ($v_{\rm eff} \ge 0$) on the SQ lattice.
Concretely, for the zero-temperature AF ($v=-1$) we have
\begin{eqnarray}
  v_{\rm eff}^{G_n}(q,-1) & \:=\: &  \Bigl( {q-1 \over q-2} \Bigr)^n \,-\, 1
    \label{eq.veff.Gn}   \\[1mm]
  v_{\rm eff}^{H_n}(q,-1)  & \:=\: &
        {q-1 \over q-2} \, \Bigl( {q-2 \over q-3} \Bigr)^{n-1} \,-\, 1
    \label{eq.veff.Hn}
\end{eqnarray}
Setting $v_{\rm eff}$ equal to the SQ-lattice ferromagnetic critical point
$v_c({\rm SQ}) = \sqrt{q}$ \cite{Baxter_book,Beffara_12},
we obtain $q_c$ for $G_n$ and $H_n$;
they have the large-$n$ asymptotic behavior
\be
   q_c(G_n)  \:\approx\:   q_c(H_n)  \;\approx\;
   {2n \over W(2n)} \,+\, O\bigl( (n/\log n)^{1/2} \bigr)
 \label{eq.qcGnHn}
\ee
where $W(x) \approx \log x - \log\log x +o(1)$
is the Lambert $W$ function \cite{note_Lambert}.
We have thus exhibited two infinite families of periodic planar lattices
on which the Potts AF has arbitrarily large $q_c$ as $n \to\infty$
\cite{note_KSS}.
These lattices are not triangulations or quadrangulations,
but they can be modified to be such and retain the phase transition,
as we now show.

\begin{figure}
  \centering
  \begin{tabular}{ccc}
  \includegraphics[width=80pt]{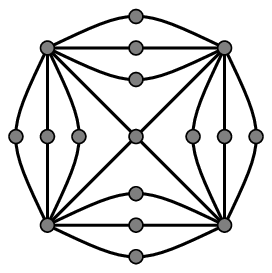} & 
  \includegraphics[width=80pt]{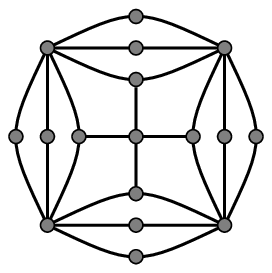} & 
  \includegraphics[width=80pt]{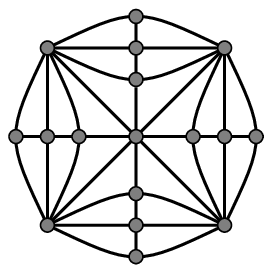} \\
  (a) & (b) & (c)  
  \end{tabular}
  \caption{%
   Unit cells of the lattices $G'_n$ (a), $G''_n$ (b) and $H'''_n$ (c) 
   for $n=3$.
   }
  \label{fig2}
\end{figure}

\paragraph{The modified lattices.}
Starting from $G_n$ or $H_n$, insert a new vertex into each octagonal face
and connect it either to the four surrounding vertices
of the original SQ lattice, to the four ``new'' vertices,
or to all eight vertices;
call these modifications ${}'$, ${}''$ and ${}'''$, respectively.
In particular, $G'_n$ and $G''_n$ are quadrangulations,
and $H'''_n$ is an Eulerian triangulation
(Fig.~\ref{fig2}).

If we integrate out the spins
at the vertices 
placed into the octagonal faces,
we obtain the model on $G_n$ or $H_n$
perturbed by a 4-spin or 8-spin interaction.
When $q$ is large, this interaction is weak (of order $1/q$)
because its Boltzmann weight is bounded between a maximum value of $q$
and a minimum value of $q-4$ or $q-8$.
We therefore expect that the new edges will have a negligible effect
on the phase transition when $q$ is large,
and that all the modified lattices will have $q_c(n)$
whose large-$n$ behavior is essentially identical to Eq.~\reff{eq.qcGnHn}.
Let us now sketch a rigorous proof \cite{KSS_in_prep} of this assertion.

\paragraph{Proof of phase transition.}
Recall first how one proves, using the Peierls argument,
the existence of ferromagnetic long-range order (FLRO)
at low temperature in the $q$-state Potts ferromagnet on the SQ lattice.
The Peierls contours are defined as the connected components
of the union of all bonds on the dual SQ lattice that separate unequal spins.
A Peierls contour $\gamma$ of length $|\gamma|$
and cyclomatic number $c(\gamma)$
comes with a weight that is
bounded above by $(q-1)^{c(\gamma)} (1+v)^{-|\gamma|}$:
here $(q-1)^{c(\gamma)}$ is a bound on the number of colorings
of the SQ lattice consistent with the contour $\gamma$.
Further, on the SQ lattice we have $c(\gamma) \le |\gamma|/2$,
and the number of contours of length $n$ surrounding a fixed site
can be bounded by $(n/2) 16^n$.
Standard Peierls reasoning
then shows that for any pair of sites $x,y$ one has
\be
   \hbox{Prob}(\sigma_x \neq \sigma_y)
   \;\le\; 
   \sum_{n=4}^\infty (n/2) 16^n (q-1)^{n/2} (1+v)^{-n}
   \,,\;
\ee
which is $\le 5/16$ whenever $1+v \ge 32 \sqrt{q}$.
This proves FLRO (the constant 32 is of course suboptimal).
The foregoing argument is valid for fixed
boundary conditions (e.g., $\sigma=1$) in the plane,
but with suitable modifications it can also be carried out
for periodic boundary conditions (i.e., on a torus).

Let us now consider the Potts antiferromagnet
on one of the six modified lattices $G'_n,\ldots,H'''_n$.
Since our goal is to show FLRO on the SQ sublattice,
we define Peierls contours exactly as we did for the SQ-lattice ferromagnet,
ignoring the spin values at all other sites.
%
%
%
Although we no longer have any simple explicit formula for the contour weights,
it is nevertheless possible to prove an upper bound
on the probability of occurrence of a contour $\gamma$
by using the technique of reflection positivity and chessboard estimates
\cite{chessboard_papers}.
Without going into details of the needed adaptations of this standard technique
for our case (see \cite{KSS_in_prep}),
we mention only that the final bound on the probability of occurrence
of a contour $\gamma$ is $(\kappa\sqrt{q-1})^{|\gamma|}$,
where $\kappa$ is the probability
that the spins on the SQ sublattice follow a fixed checkerboard pattern
(say, 1 on the even sublattice and 2 on the odd sublattice)
raised to the power 1/volume.
This latter probability is easy to bound explicitly,
yielding $\kappa \le [1+ v_{\rm eff}(q,v)]^{-1} [q/(q-5)]^{1/2}$,
where $v_{\rm eff}(q,v)$ is the one for the corresponding unmodified lattice
$G_n$ or $H_n$.
This implies that, for all the lattices $G'_n,\ldots,H'''_n$,
there is FLRO on the SQ sublattice whenever
$6 \le q \le q_c(G_n) - O(1)$
[cf.\ Eq.~\reff{eq.qcGnHn}]
and $v$ is close to $-1$ (low temperature).

Let us also remark that the lattice $G''_2$
is covered by the general theory of \cite{Kotecky-Sokal-Swart},
where it is proven that $q_c > 3$;
moreover, a minor modification proves the same result
for $G''_n$ for all $n \ge 2$.

\begin{figure}
\begin{center}
\includegraphics[width=0.5\columnwidth]{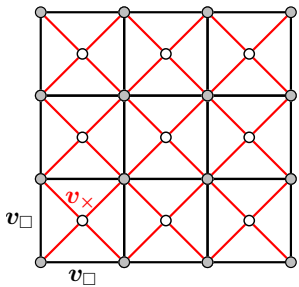}
\end{center}
\vspace*{-5mm}
\caption{
   (Color online) The union-jack (UJ) lattice.
}
\label{fig_UJ_pic}
\end{figure}

\begin{figure}[p]
\begin{center}
\includegraphics[width=0.75\columnwidth]{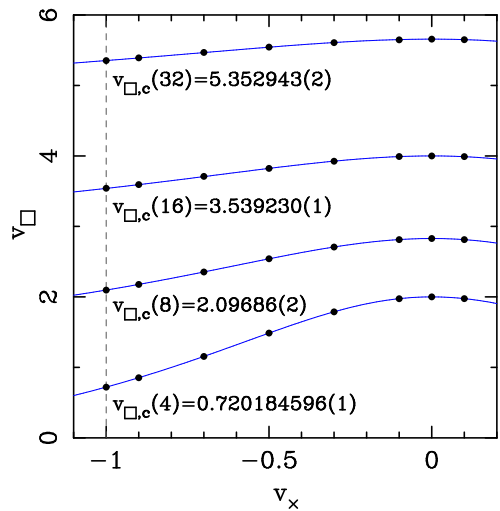}
\end{center}
\vspace*{-6mm}
\caption{
   (Color online)  Estimated phase boundaries for the $q=4,8,16,32$
   Potts models on the union-jack (UJ) lattice,
   from the Jacobsen--Scullard method
   (blue curve and numerical values of $v_{\Box,{\rm c}}$ at $v_\times=-1$)
   and Monte Carlo simulations (black points).
}
\vspace*{3mm}
\label{fig_UJ}
\end{figure}

\begin{table}[p]
\footnotesize
\begin{center}
\begin{tabular}{|r||r|r|r|r||r|}
\hline
$\;n\;$ &  \multicolumn{1}{|c|}{$q_c(G_n)$} &  \multicolumn{1}{|c|}{$q_c(G'_n)$}
     &  \multicolumn{1}{|c|}{$q_c(H_n)$} &  \multicolumn{1}{|c||}{$q_c(H'_n)$}
     &  \multicolumn{1}{|c|}{$2n/W(2n)$}   \\
     &  \multicolumn{1}{|c|}{(exact)}    &  \multicolumn{1}{|c|}{(JS)}
     &  \multicolumn{1}{|c|}{(exact)}    &  \multicolumn{1}{|c||}{(JS)}
     &  \multicolumn{1}{|c|}{(asymp.)}    \\
\hline
1    &   2.618034   &  3.74583(8)\x &    2.618034  & 3.74583(8)\x 
                                         &  2.345751   \\
2    &   3.448678   &  4.48805(4)\x &  4\hphantom{.000000}  &  4.80794(5)\x  
                                         &  3.327322   \\
4    &   4.942152   &  5.87902(5)\x &  5.617069  &  6.39269(4)\x  
                                         &  4.981903   \\
8    &   7.565625   &  8.40372(3)\x &  8.304127  &  9.04238(2)\x  
                                         &  7.792741   \\
16   &  12.164794   & 12.91503(1)\x & 12.939420  & 13.63221(2)\x   
                                         & 12.621338   \\
32   &  20.270897   & 20.945341(3)  & 21.068717  & 21.711603(3)   
                                         & 21.016077   \\
64   &  34.667189   & 35.276721(3)  & 35.482095  & 36.074775(3)   
                                         & 35.780223   \\
\hline
\end{tabular}
\end{center}
\vspace*{-3mm}
\caption{
   Estimates of $q_c(n)$ for the lattices $G_n, G'_n, H_n, H'_n$
   from the Jacobsen--Scullard (JS) method or the exact solution,
   and their large-$n$ asymptote $2n/W(2n)$ from Eq.~\reff{eq.qcGnHn}.
}
\label{table_qcn}
\end{table}

\begin{figure}[p]
\vspace*{1mm}
\begin{center}
\includegraphics[width=0.8\columnwidth]{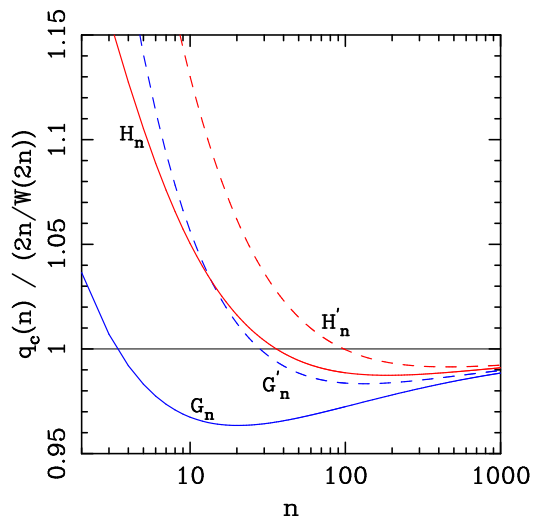}
\end{center}
\vspace*{-5mm}
\caption{
   (Color online)  Estimates of $q_c(n)$ for the lattices $G_n, G'_n, H_n, H'_n$
   divided by their large-$n$ asymptote $2n/W(2n)$.
}
\label{fig_qcn}
\end{figure}

\paragraph{Data for lattices $G'_n$ and $H'_n$.}
The lattices $G'_n$ and $H'_n$ for all $n$ can be reduced
to the union-jack (UJ) lattice (Fig.~\ref{fig_UJ_pic}) with
$v_\times = v$ and $v_\Box =$ a suitable $v_{\rm eff}(q,v)$
[cf.\ Eqns.~\reff{eq.veff.Gn}/\reff{eq.veff.Hn} when $v=-1$];
of course the same reduction holds for $G_n$ and $H_n$
by setting $v_\times = 0$.
We obtained high-precision estimates of the phase boundary
of the UJ model in the $(v_\times, v_\Box)$-plane
by using the Jacobsen--Scullard (JS) method \cite{JS1}
with untwisted square bases of size up to $7 \times 7$ (294~edges) \cite{JS2}.
We checked these results for $q=4,8,16,32$
by Monte Carlo simulations using a cluster algorithm \cite{note_algorithm}.
The estimated phase boundaries from both methods
are shown in Fig.~\ref{fig_UJ},
along with the numerical estimates of $v_{\Box,{\rm c}}$ at $v_\times = -1$
from the JS method.
The estimates of $q_c(n)$ for the lattices $G_n, G'_n, H_n, H'_n$
from the JS method (or the exact solution) are shown in Table~\ref{table_qcn},
where they are compared with the predicted large-$n$ asymptote
$q_c(n) \approx 2n/W(2n)$ from Eq.~\reff{eq.qcGnHn}.
The functions $q_c(n)$ divided by $2n/W(2n)$
are plotted in Fig.~\ref{fig_qcn}.
Note that $q_c(G'_n) > q_c(G_n)$ and $q_c(H'_n) > q_c(H_n)$,
   in accordance with the intuitive idea that the AF edges
   associated to the modification $\,{}'\,$
   {\em enhance}\/ the ferromagnetic ordering on the SQ sublattice
   \cite{note_corrineq}.

\begin{figure}
\begin{center}
\includegraphics[width=0.7\columnwidth]{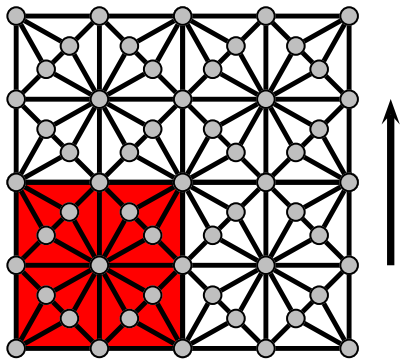} 
\end{center}
\vspace*{-6mm}
\caption{
   (Color online) The lattice $H'''_2$
       (rotated $45^\circ$ from Fig.~\ref{fig2})
       with $L=2$;
       a unit cell is shown in red,
       and the transfer direction is indicated with an arrow.
}
\label{fig_TM_pic}
\end{figure}

\paragraph{Data for lattices $G''_n$ and $H'''_n$.}
We studied the lattices $G''_n$ and $H'''_n$ for $n=1,2,4,8,16,32,64$
(note that $G''_1 =$ SQ \cite{Salas_98} and $H'''_1 =$ UJ \cite{union-jack})
at $v=-1$,
using transfer matrices with cylindrical boundary conditions
on widths $L = 1,2,3,4$ unit cells (Fig.~\ref{fig_TM_pic}).
The computational complexity is linear in $n$.
We estimated the location of the phase transition
(which we expect to be first-order whenever $q_c > 4$)
using the crossings of the energies $E_L(q)$ \cite{Borgs_91}:
the results are shown in Table~\ref{table_TM}.

\begin{figure}
\vspace*{2mm}
\begin{center}
\includegraphics[width=0.8\columnwidth]{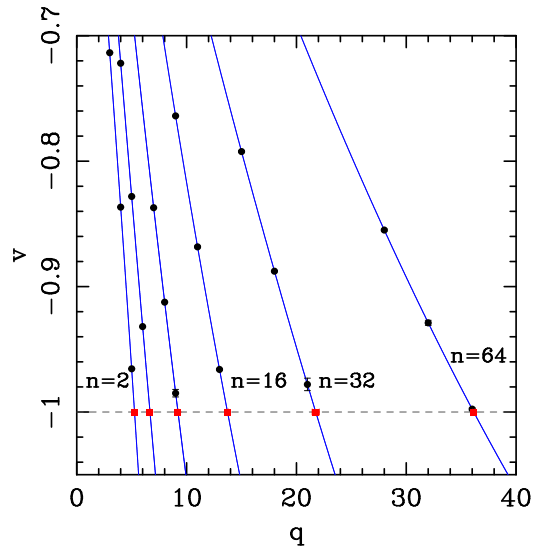}  
\end{center}
\vspace*{-5mm}
\caption{
   (Color online)  Monte Carlo estimates of $v_c$ for the lattices $H'''_n$
       (black points) and their quadratic fit (blue curves),
       together with the extrapolated values $q_c$ (red squares).
}
\label{fig_TM}
\end{figure}

For $H'''_n$ we checked these results by Monte Carlo:
for three integer values of $q$ below the estimated~$q_c$
we simulated the model at finite temperature
and estimated the transition point $v_c(q)$;
we then performed linear and quadratic extrapolations
to locate the point~$q_c$ where $v_c = -1$.
The results are shown in Table~\ref{table_TM} and Fig.~\ref{fig_TM}
and agree well with the transfer-matrix estimates.
For $q \gtapprox 8$ the specific heat diverges at the transition point
like $L^{\approx 2}$, in agreement with the finite-size-scaling prediction
for a first-order transition;
for $4 < q \ltapprox 8$ the transition is presumably also first-order
but with a large correlation length $\xi$, so that we are unable to observe
the true $L \gg \xi$ asymptotic behavior.

\paragraph{Conclusion.}
When a 2D model admits a height representation,
it must be either critical or ordered.
Until now criticality seemed to be the most common case,
even though examples of order were known.
But here we have exhibited several infinite families of 2D lattices
--- some of which are quadrangulations or Eulerian triangulations ---
in which the Potts antiferromagnet admitting a height representation
($q=3$ or 4, respectively) is not only ordered
but is in fact ``arbitrarily strongly ordered''
in the sense that $q_c$ is arbitrarily large.
This unexpected result suggests that the prior belief
may have things precisely backwards.
Perhaps criticality is an exceptional case
--- arising, for instance, in situations with special symmetries ---
and order is to be generically expected.
A key open question raised by this work
is to understand why criticality arises when it does.

\begin{table}
\vspace{3mm}
\begin{center}
\begin{tabular}{|r||r||r|r||r|}
\hline
$\;n\;$  &  \multicolumn{1}{|c||}{$\:q_c(G''_n)\:$}
     &  \multicolumn{1}{|c|}{$\:q_c(H'''_n)\:$}
     &  \multicolumn{1}{|c||}{$\:q_c(H'''_n)\:$}
     &  \multicolumn{1}{|c|}{$\,2n/W(2n)\,$}   \\
     &  \multicolumn{1}{|c||}{(TM)}
     &  \multicolumn{1}{|c|}{(TM)}
     &  \multicolumn{1}{|c||}{(MC)}
     &  \multicolumn{1}{|c|}{(asymp.)}    \\
\hline
1    &  3\hphantom{.00(0)} &   4.31(3)    &              &  2.345751$\:$    \\
2    &  3.63(2)            &   5.27(1)    &   5.26(2)\x  &  3.327322$\:$    \\
4    &  5.02(1)            &   6.68(1)    &   6.67(3)\x  &  4.981903$\:$    \\
8    &  7.60(1)            &   9.21(1)    &   9.21(7)\x  &  7.792741$\:$    \\
16   & 12.18(2)            &  13.73(2)    &  13.73(10)   & 12.621338$\:$    \\
32   & 20.29(3)            &  21.76(3)    &  21.76(32)   & 21.016077$\:$    \\
64   & 34.70(5)            &  36.10(5)    &  36.14(8)\x  & 35.780223$\:$    \\
\hline
\end{tabular} \\[5mm]
\end{center}
\vspace*{-5.5mm}
\caption{
   Estimates of $q_c$ for the lattices $G''_n$ and $H'''_n$
   from transfer matrices (TM) and Monte Carlo (MC),
   and their large-$n$ asymptote from Eq.~\reff{eq.qcGnHn}.
}
\label{table_TM}
\end{table}

\begin{acknowledgments}
This work was supported in part by
NSFC grants 10975127 and 11275185, the Chinese Academy of Sciences,
French grant ANR-10-BLAN-0414, the Institut Universitaire de France,
Spanish MEC grants FPA2009-08785 and MTM2011-24097,
Czech GA\v CR grant P201/12/2613,
US NSF grant PHY--0424082,
and a computer donation from the Dell Corporation.
\end{acknowledgments}


\begin{thebibliography}{99}
\bibitem{note_TRI_q=2}  On the triangular lattice, both $q=2$ and $q=4$ 
are critical at zero temperature and have height representations
\cite{Blote_82_etc} \cite{Moore_00}, but $q_c = 4$.
\bibitem{note_KSS}  A more complicated example with this property
was given in the last paragraph of \cite{Kotecky-Salas-Sokal}.
\end{thebibliography}
\end{document}